\documentclass[lettersize,journal]{IEEEtran}
\usepackage{amsmath,amsfonts}
\usepackage{float}
\usepackage{algorithmic}
\usepackage{algorithm}
\usepackage{array}
\usepackage[caption=false,font=normalsize,labelfont=sf,textfont=sf]{subfig}
\usepackage{textcomp}
\usepackage{stfloats}
\usepackage{url}
\usepackage{verbatim}
\usepackage{graphicx}
\usepackage{cite}
\newtheorem{theorem}{Theorem}

\begin{document}

\title{Low-Complexity Decoding for Low-Rate Block Codes of Short Length Based on Concatenated Coding Structure}

\author{Mao-Chao Lin,~\IEEEmembership{Member,~IEEE,} Shih-Kai Lee,~\IEEEmembership{Member,~IEEE,} Pin Lin, Ching-Chang Lin, \\ Chia-Chun Chen,  Teng-Yuan Syu,  Huang-Chang Lee,~\IEEEmembership{Member,~IEEE,}
\thanks{This work was financially supported by the Ministry of Science and Technology of Taiwan under Grants NSTC 112-2221-E-002 -169.}
\thanks{M.-C. Lin is with the Department of Electrical Engineering, National Taiwan University, Taipei 10617, Taiwan, R.O.C. (e-mail: mclin@ntu.edu.tw)}
\thanks{S.-K. Lee is with the Department of Electrical Engineering, Yuan Ze University, Taoyuan City 32003, Taiwan, R.O.C. (e-mail: sklee@saturn.yzu.edu.tw)}
\thanks{P. Lin is with Synology Inc., New Taipei City, Taiwan, R.O.C. (e-mail: r09942129@g.ntu.edu.tw)}
\thanks{C.-C Lin and T.-Y Syu are with the Graduate Institute of Communication Engineering, National Taiwan University, Taipei 10617, Taiwan, R.O.C. (e-mail: r11942139@ntu.edu.tw, r12942069@ntu.edu.tw)}
\thanks{C.-C. Chen is with Realtek Semiconductor Corp., Hsinchu, Taiwan, R.O.C. (e-mail: ppgame2010231@gmail.com)}
\thanks{H.-C. Lee is with the Department of Electrical Engineering, Taiwan Ocean University, Keelung City 202301, Taiwan. R.O.C. (e-mail: hclee@mail.ntou.edu.tw)}
}

\markboth{Journal of \LaTeX\ Class Files,~Vol.~14, No.~8, August~2021}%
{Shell \MakeLowercase{\textit{et al.}}: A Sample Article Using IEEEtran.cls for IEEE Journals}


\maketitle

\begin{abstract}
To decode a short linear block code, ordered statics decoding (OSD) and/or the $A^*$ decoding are usually considered.  Either OSD or the $A^*$ decoding utilizes the magnitudes of the received symbols to establish the most reliable and independent positions (MRIP) frame. A restricted searched space can be employed to achieve near-optimum decoding with reduced decoding complexity.  For a low-rate code with large minimum distance, the restricted search space is still very huge.

 We propose to use concatenated coding to further restrict the search space by proposing an improved MRIP frame.  The improved MRIP frame is founded according to magnitudes of log likelihood ratios (LLRs) obtained by the soft-in soft-out (SISO) decoder for the inner code.  

We focus on the construction and decoding of several $(n,k)$ = (128,36) binary linear block codes based on concatenated coding. We use the (128,36) extended BCH (eBCH) code as a benchmark for comparison.   
Simulation shows that there exist constructed concatenated codes which are much more efficient than the (128,36) eBCH code.  Some other codes of length 128 or close to 128 are also constructed to demonstrate the efficiency of the proposed scheme.

\end{abstract}


\section{Introduction}
\IEEEPARstart{I}{n} the 5G communication systems, services with ultra-reliable and low latency communications (URLCC) \cite{8594709} are desired. If the decoding time is negligible, then error-correcting codes with short block lengths can meet the demand of low latency. Error-correcting codes with low coding rates can help to meet the demand of ultra reliability. 

For binary linear $(n,k,d_{min})$ block codes with code length $n$ no greater than 200, message passing algorithm (MPA) is not efficient, where $k$ is the number of message bits and $d_{min}$ is the minimum distance.  
According to our knowledge, order statistics decoding (OSD) \cite{412683} and $A^*$ decoding \cite{259636}\cite{7182759}which is also known as priority-first search decoding \cite{259636} are most competitive for decoding short linear block codes.  In this research, we take the $A^*$ decoding approach. However, our concept can also be applied to the OSD approach.

Both the $A^*$ decoding and the OSD employ the property of most reliable and independent positions (MRIP) or equivalently the property of most reliable basis (MRB) by which the received symbols are permuted so that the associated generator matrix is systematic and the $k$ most reliable and independent symbols correspond to the $k$ message bits.   
Simplified decoding obtained by restricting the search space to no more than $\lambda$ errors in the $k$ MRIP symbols has been proposed in \cite{412683} and is referred to as OSD-$\lambda$ here.  In OSD-$\lambda$, the number of candidates in the search space is restricted to at most $\binom{k}{\lambda}+\binom{k}{\lambda-1} +\cdots+\binom{k}{0}$.  In \cite{412683}, it has been shown that OSD-$\lambda$ is asymptotically optimum if $\lambda \ge \min\{\lceil d_{min}/4 \rceil-1,k\}$. 

  There are many further works related to OSD-$\lambda$ such as \cite{1291728}, \cite{4039664} and \cite{1576590}.  In any case, the parameter $\lambda$ is essential. In \cite{6952162}, the concept of OSD-$\lambda$ has been incorporated into $A^*$ decoding and is referred to as PC-$\lambda$ (path constraint-$\lambda$).  A further simplified version of PC-$\lambda$ is called PC-out-$\lambda$ \cite{8664355}, which is implemented by requiring each searched partial path with $\lambda$ errors deviated from the hard decision values of the MRIP symbols to be extended directly to the complete codeword path.

In this paper, we study the the construction and decoding of low-rate codes.  In particular, we focus on the $(n,k)$ = (128,36) binary linear block codes. We use the (128,36,32) extended BCH (eBCH) code as a benchmark for comparison.  The value of $\min\{\lceil 32/4 \rceil-1,k\}$ =7 is so high that unbearably high decoding complexity will be required if $\lambda$ = 7 is used in either OSD-$\lambda$ or PC-out-$\lambda$.    
In \cite{8594709}, $\lambda$ = 5 is adopted in OSD-$\lambda$ for the (128,36) eBCH code.  To beat the (128,36,32) eBCH code, we construct several (128,36) concatenated codes by using a (16,9) Reed-Solomon code over GF($2^4$) as the outer code and specially chosen binary codes with rate 1/2 as inner codes.  These specially chosen binary inner codes include the (8,4,4) extended Hamming code, a (16,8,5) code \cite{192220}, (2,1,4) and (2,1,6) binary convolutional code \cite{slin2004} respectively, where a $(n,k,m)$ binary convolutional code with memory order $m$ takes $k$ input bits and generates $n$ output bits for each time unit.

The concatenated coding structure has the advantage that we can employ the inner code to obtain improved MRIP which is more reliable than the conventional MRIP.  With the more reliable MRIP, we can implement the $A^*$ decoding with lower $\lambda$ value, i.e., lower decoding complexity while the error performances are not sacrificed and may be better in some cases.  The advantage of the proposed (128,36) concatenated codes over the (128,36) eBCH code can be very significant.  In particular, the concatenated code using the (2,1,6) convolutional code has the best performances.

To see the efficiency of the concatenated coding for other short block codes, we also construct a (130,65) concatenated code and a (128,24) concatenated code for comparison with the (128,64) eBCH code and the (128,22) eBCH code.  We will see that the advantage of the (130,65) concatenated code over the (128,64) eBCH code is not significant.   This result suggests that the concatenated coding is efficient for applications with rate less than 1/2. 

In the conventional $A^*$ decoding, a stack with its contents ordered according to the values of their metrics is used.  In this paper, we use a modified stack arrangement for which the ordering is not needed.  The first advantage is that the huge effort of comparisons for ordering is no longer needed.  The second advantage is that its capability of searching all the goal nodes with distance of $i$ from the hard decision vector at the MRIP positions before those with distance of $j$, where $i < j$.  The disadvantage is that a large stack is needed.  Since no ordering is needed, the requirement of a large stack size is not a problem.

Recently, polar adjusted convolutional (PAC) code has been designed for short code applications \cite{arikan2019seq}\cite{zhu23}.  However, according to our knowledge, for short codes with low-code rates the code performances are quite inferior to our comparable codes.  For example for $E_b/N_0$ = 3 dB, in \cite{zhu23}, a (128,32) PAC code is designed for which its dispersion bound is above $10^{-4}$ while our constructed (128,36) concatenated code can achieve BLER about $3.5 \times 10^{-5}$. 

The presentation of this paper is organized as follows.  In Section II, we review the $A^{*}$ algorithm.  In Section III, the modified stack arrangement without ordering is presented.  Section IV shows the proposed binary codes based on concatenated coding structure together with simulated block error rate (BLER) performances.  Concluding Remarks are provided in Section V.

\section{Review for the Basic OSD and $A^*$ algorithm}
We here review the basic OSD and $A^*$ decoding algorithm.   Both decoding methods share the concept of most reliable and independent positions (MRIP).
\subsection{MRIP}
Let $C$ be an $(n,k)$ binary block code with generator matrix ${G}$. Let ${{\bf c}}$ = $({c}_0, {c}_1,\cdots, {c}_{n-1})$ be a codeword of $C$.  Let ${{\bf r}}$ = $({r}_0, {r}_1,\cdots, {r}_{n-1})$ be the received vector, where $r_i = (-1)^{c_{i}}$ + $w$ and $w$ is the additive Gaussian noise with variance $\sigma_{w}^2$.  The MRIP frame is implemented according to the magnitudes of the symbols in ${\bf r}$.  The first step of achieving the MRIP frame for decoding is applying permutation $\Pi_{1}$ to ${{\bf r}}$ to obtain ${\bf r}'$ = $\Pi_{1}({\bf r})$ = $(r'_0, r'_1,\cdots, r'_{n-1})$ such that $|r'_i| \ge |r'_j|$ for $i < j$.  Applying the same permutation to the columns of ${G}$, we have the generator matrix $G' = \Pi_{1}(G)$.  The first $k$ columns of $G'$ may not be linearly independent.  Permute the first $k$ linearly independent columns as the first $k$ columns of a generator matrix $G"$ = $\Pi_{2}\Pi_{1}(G)$, which can then be converted into a systematic generator matrix $\hat{G}$, where the first $k$ columns is a $k \times k$ identity matrix.  With the same permutation, we permute ${\bf r}'$ into $\hat{\bf r}$ = $\Pi_{2}({\bf r}')$ = $(\hat{r}_0, \hat{r}_1,\cdots, \hat{r}_{n-1})$, where $|\hat{r}_i| \ge |\hat{r}_j|$ for $0 \le i < j \le k-1$. 

 According to $\hat{G}$, we can generate a code space of $2^k$ codewords of $\hat{C}$, of which each codeword is $\hat{\bf c}$ = $(\hat{c}_0, \hat{c}_1, \cdots, \hat{c}_{n-1})$  = $(\hat{c}_0, \hat{c}_1, \cdots$  $\hat{c}_{k-1})\hat{G}$, where $\hat{c}_0, \hat{c}_1, \cdots$  $\hat{c}_{k-1}$ represent the $k$ message bits.  Each codeword of $\hat{C}$ is a permuted version of a codeword of $C$, i.e., $\hat{\bf c}$ = $\Pi_{2}\Pi_{1}({\bf c})$.

Let $\hat{\bf z}$ = $(\hat{z}_0, \hat{z}_1,\cdots, \hat{z}_{n-1})$ be the hard decision of $\hat{\bf r}$, where $\hat{z}_j = 0$ if $\mbox{sign}(\hat{r}_j)$ = $1$ and $\hat{z}_j$ = $1$ if $\mbox{sign}(\hat{r}_j)$ = $-1$. The metric associated to a code bit $\hat{c}_j$ is $M(\hat{c}_j)= 0$ if  $\hat{c}_j$  = $\hat{z}_j$ .  If  $\hat{c}_j \neq \hat{z}_j$, we have $M(\hat{c}_j)= |\hat{r}_j|$.  The correlation discrepancy between $\hat{\bf r}$ and $\hat{\bf c}$ is defined as
\begin{equation}
M(\hat{\bf r},\hat{\bf c}) = \sum_{j=0}^{n-1} M(\hat{c}_j).
\end{equation}
\subsection{OSD with order $\lambda$}
Let ${\bf x}_{a}^{b}$ = $(x_a, x_{a+1}, x_{a+2}, ..., x_{b})$ denote a subvector of a vector ${\bf x}$ = $(x_0, x_{1}, x_{2}, ..., x_{n-1})$. Then, for $0 \le l \le k-1$, the Hamming distance between $(\hat{c}_0, \hat{c}_1, ..., \hat{c}_{l})$ and $(\hat{z}_0, \hat{z}_1, ..., \hat{z}_{l})$, i.e., $d_{H}$($(\hat{c}_0$, $\hat{c}_1$, ..., $\hat{c}_{l})$, $(\hat{z}_0, \hat{z}_1, ..., \hat{z}_{l})$), can be represented as $d_{H}(\hat{\bf c}_{0}^{l}, \hat{\bf z}_{0}^{l})$.

 For $1 \le j \le \lambda$, compute $M(\hat{\bf r},\hat{\bf c})$ for each $\hat{\bf c}$ of which the associated $d_{H}(\hat{\bf c}_{0}^{k-1}, \hat{\bf z}_{0}^{k-1})$ = $j$.   The $\hat{\bf c}$ with the smallest $M(\hat{\bf r},\hat{\bf c})$ is the desired one.  Applying the reverse permutation, we have the decoded codeword $(\Pi_{2}\Pi_{1})^{-1}\hat{\bf c}$ = ${\bf c}$.  The number of candidates in the search space of OSD with order $\lambda$, abbreviated as OSD-$\lambda$ is at most $\binom{k}{\lambda}+\binom{k}{\lambda-1} +\cdots+\binom{k}{0}$.

\subsection{$A^*$ decoding}
The original $A^*$ decoding is similar to OSD with order $k$ except that the set of all the codewords is arranged as a code tree of which the priority search is employed.

We generate a code tree, of which a path represents a codeword of $\hat{C}$, i.e., $\hat{\bf c}$ = $(\hat{c}_0, \hat{c}_1, \cdots, \hat{c}_{n-1})$  = $\hat{\bf c}_{0}^{k-1}\hat{G}$.  
 In this tree, each path stems from a root node at level 0. For $0 \le i \le k-1$, two edges labelled by $\hat{c}_i = 0$ and $\hat{c}_i = 1$ respectively will be extended from a node at level $i$.   For $i=k$, the complete path (codeword of $\hat{C}$) $(\hat{c}_0$, $\hat{c}_1$, $\cdots$, $\hat{c}_{k-1}$, $\cdots$, $\hat{c}_{n-1})$ is determined by the message part $\hat{\bf c}_{0}^{k-1}$ = $(\hat{c}_0, \hat{c}_1, \cdots$ $\hat{c}_{k-1})$.  

The $A^*$ decoding executes the priority search using the code tree together with a stack.  The priority search begins from the root node (node 0) of a search tree.  Set the path metric $PM(0)$ = 0. Set the best metric $M_{best}$ = $\infty$.
\begin{enumerate}
  \item 
Node 0 is extended to node 1 and node 2 respectively.  
\begin{itemize}
\item %
 Node 1 represents the path $PH(1)$ = $\{\hat{c}_0 = 0\}$ with path metric $PM (1)$ = $M(\hat{c}_0=0)$.  Node 2 represents the path $PH(2)$ = $\{\hat{c}_0 = 1\}$ with path metric $PM(2)$ = $M(\hat{c}_0=1)$. 
\item %
 Node 1 and node 2 are placed in a stack, where the node with the smaller metric are placed on the top.
 \end{itemize}
  \item 
 Suppose node $j$ is the one at the top of the stack, where its path is $PH(j)$ = $(\hat{c}_0, \hat{c}_1, ..., \hat{c}_{l-1})$ and its metric is $PM(j)$, where $0 < l \le k-2$.  We extend node $j$ to node $j+1$ and node $j+2$ respectively.
 \begin{itemize}
 \item %
 Node $j+1$ represents the path $PH(j+1)$ = $(\hat{c}_0, \hat{c}_1$, ..., $\hat{c}_{l-1}, \hat{c}_{l} = 0)$ with path metric $PM(j+1)$ = $PM(j) + M(\hat{c}_l = 0)$. Node $j+2$ represents the path $PH(j+2)$ = $(\hat{c}_0, \hat{c}_1$, ..., $\hat{c}_{l-1}, \hat{c}_{l} = 1)$ with path metric $PM(j+2)$ = $PM(j) + M(\hat{c}_l = 1)$. 
 \item %
 One of node $j+1$ and node $j+2$ has path metric $PM(j)$, while the other node has path metric $PM(j)+|\hat{r}_{j+1}|$.  The node with metric $PM(j)$ is placed at the top of the stack.  If $PM(j)+|\hat{r}_{j+1}| < M_{best}$, the node with path metric $PM(j)+|\hat{r}_{j+1}|$ is inserted into the stack so that the nodes in the stack are in ascending order according to the path metrics. Otherwise the node with path metric $PM(j)+|\hat{r}_{j+1}|$ will be deleted.
 \end{itemize}
 \item 
Suppose node $j$ is the one at the top of the stack, where its path is $PH(j)$ = $(\hat{c}_0, \hat{c}_1$, $...$, $\hat{c}_{k-2})$ and its metric is $PM(j)$.  We extend node $j$ to two goal nodes, which are node $j+1$ and node $j+2$ with the complete code paths (i.e., codewords) $PH(j+1)$ = $(\hat{c}_0, \hat{c}_1$, ..., $\hat{c}_{k-2}, \hat{c}_{k-1} = 0)\hat{G}$ and $PH(j+2)$ = $(\hat{c}_0, \hat{c}_1$, ..., $\hat{c}_{k-2}, \hat{c}_{k-1} = 1)\hat{G}$ respectively.   The associated path metrics are $PM(j+1)$ and $PM(j+2)$ respectively, where each path metric is obtained by the summation of the bit metrics of code bits in the codeword.
\begin{itemize}
\item 
i) If $\min\{PM(j+1),PM(j+2)\}$ $\ge M_{best}$, then both node $j+1$ and node $j+2$ are deleted.
\item 
ii) If $\min\{PM(j+1),PM(j+2)\}$  $< M_{best}$, then the one between node $j+1$ and node $j+2$ with the smaller path metric is set as $\hat{\bf c}_{best}$.   Set $M_{best}$ = $\min\{PM(j+1),PM(j+2)\}$. Delete the other node and delete any node in the stack for which the associated path metric is greater than $M_{best}$.
\end{itemize}
\item 
Repeat the procedure of steps 2) and 3) until the stack is empty.  The current $\hat{\bf c}_{best}$ is the desired codeword $\hat{\bf c}$ of $\hat{C}$.  Applying the reverse permutation, we have the decoded codeword $(\Pi_{2}\Pi_{1})^{-1}\tilde{\bf c}$ = ${\bf c}$ of $C$.
\end{enumerate}

\subsection{PC-$\lambda$ and PC-out-$\lambda$}
 In \cite{6952162}, the concept of OSD-$\lambda$ has been incorporated into $A^*$ decoding and is referred to as PC-$\lambda$ (path constraint-$\lambda$).  This is accomplished by modifying step 2) of the $A^*$ decoding by adding the constraint, "If $d_{H}(\hat{\bf c}_{0}^{l}, \hat{\bf z}_{0}^{l}) > \lambda$, then the associated node is deleted."

PC-$\lambda$ can be further modified to PC-out-$\lambda$ \cite{8664355}.  The added constraint is modified as
"If $d_{H}(\hat{\bf c}_{0}^{l}, \hat{\bf z}_{0}^{l}) = \lambda$, then this node $\hat{\bf c}_{0}^{l}$ is directly extended to the goal node $(\hat{c}_0, \hat{c}_1$, ..., $\hat{c}_{l}$, $\hat{z}_{l+1}, ..., \hat{z}_{k-1})\hat{G}$."

\subsection{Stopping Criterion}
In OSD-$\lambda$, the candidate codeword $\hat{\bf c}$ in the search space is tested according to $d_{H}(\hat{\bf c}_{0}^{k-1}, \hat{\bf z}_{0}^{k-1})$.  Moreover, a codeword $\hat{\bf c}$ with a smaller $d_{H}(\hat{\bf c}_{0}^{k-1}, \hat{\bf z}_{0}^{k-1})$ is tested before any other codeword $\hat{\bf c}'$ with a larger $d_{H}(\hat{\bf c'}_{0}^{k-1}, \hat{\bf z}_{0}^{k-1})$ so as to reduce the decoding complexity.  

An early stopping criterion \cite{412683} can identify whether a tested candidate codeword $\hat{\bf c}$ is the maximum likelihood codeword so that unnecessary tests can be avoided.  Define 
\begin{equation}
D(\hat{\bf c}) = \{l_1, l_2, ..., \} = \{j : \hat{c}_j = \hat{z}_j \ \mbox{with} \ 0 \le j < n\},
\end{equation}
where $|\hat{r}_{l_i}| < |\hat{r}_{l_j}|$ for $i < j$.
 Let $q$ = $d_{min}$ - $d_{H}(\hat{\bf c},\hat{\bf z})$ and let
\begin{equation}
D_{q}(\hat{\bf c}) = \{l_1, l_2, ..., l_q\} \subseteq D(\hat{\bf c}).
\end{equation}
Define a threshold
\begin{equation}
M_{TH,\hat{\bf c}} = \sum_{j \in D_{q}(\hat{\bf c})} |\hat{r}_j|.
\end{equation}
It has been shown in \cite{412683}\cite{61132} that $\hat{\bf c}$ is the maximum likelihood codeword for $\hat{\bf r}$ if $M(\hat{\bf r}, \hat{\bf c}) \le M_{TH,\hat{\bf c}}$.   

The $A^*$ algorithm can also utilize the stopping criterion for early termination of the decoding by testing whether a $M(\hat{\bf r}, \hat{\bf c}_{best})$ is no greater than $M_{TH,\hat{\bf c}_{best}}$ for $\hat{\bf c}_{best}$. 

Calculation of the threshold $M_{TH,\hat{\bf c}}$  requires the knowledge of the code minimum distance $d_{min}$.  In this paper, we propose a threshold ${M}_{TH,\alpha}$ that is not codeword dependent and does not need the information of $d_{min}$, where 
\begin{equation}
M_{TH,\alpha} = \alpha\sum_{j=0}^{n-1} |\hat{r}_j|.
\end{equation} 

Note that, $\hat{\bf c}$ is not necessarily the maximum likelihood codeword for $\hat{\bf r}$ if $M(\hat{\bf r}, \hat{\bf c}) \le M_{TH,\alpha}$.  
Let $P(\alpha)$ be the probability that $\hat{\bf c}$ is not the maximum likelihood codeword for $\hat{\bf r}$ when $M(\hat{\bf r}, \hat{\bf c}) \le M_{TH,\alpha}$.
We can reduce the block error rate (BLER) by choosing $\alpha$ such that $P(\alpha)$ is very small.   In practice, we usually choose $\alpha$ to be 0.045 or 0.05 which is obtained through simulation.  Simulations implemented for some codes show that the performances (including error rates and complexities) of using either $M_{TH,\alpha}$ and $M_{TH,\hat{\bf c}}$ are usually very similar. In case $d_{min}$ is not available, using $M_{TH,\alpha}$ is a good option.

\section{Stack Design for $A^*$ Decoding}
The stack design will affect the performances of the $A^*$ decoding, where the performances include the BLER and the decoding complexity.  There are two measures for evaluating the decoding complexity of the $A^*$ decoding.  One is the number searched edges per message bit.  The other is the number of comparisons per message bit.  
\subsection{Conventional Stack in Ascending Order} For the conventional stack used in the $A^*$ decoding, the nodes needed to be arranged such that the associated path metrics are in ascended order.  If a new node is to be inserted into such a stack with $N$ nodes, the number of comparisons needed is $O(\log_{2}N)$. This is the major barrier in applying a large stack in the $A^*$ decoding.  

\subsection{Modified Stack without Ordering}
We now propose a stacking method which essentially avoids the need of comparison.  The only modification on the conventional stack is as follows.  
\begin{itemize}
\item %
The last two sentences of Step 2) of Section II.C is now modified as "If $PM(j)+|\hat{r}_{j+1}| < M_{best}$, the node with path metric $PM(j)+|\hat{r}_{j+1}|$ is placed at the bottom of the stack. Otherwise the node with path metric $PM(j)+|\hat{r}_{j+1}|$ will be deleted."
\end{itemize}
 The modified stack is not in ascending order now.

This modified stack design has two advantages :
\begin{enumerate}
\item 
There is no need for finding a position in order to insert a node in the stack while keeping the path metrics in ascending order.  
\item  
In OSD-$\lambda$, the candidate codeword $\hat{\bf c}$ in the search space is tested according to $d_{H}(\hat{\bf c}_{0}^{k-1},\hat{\bf z}_{0}^{k-1})$, where a codeword $\hat{\bf c}$ with a smaller $d_{H}(\hat{\bf c}_{0}^{k-1},\hat{\bf z}_{0}^{k-1})$ is tested before any other codeword $\hat{\bf c}'$ with a larger $d_{H}(\hat{\bf c'}_{0}^{k-1},\hat{\bf z}_{0}^{k-1})$.     The operation of the modified stack has similar property which is stated in the following Theorem.
\end{enumerate}

\begin{theorem}
For the operation of the modified stack, a path $\hat{\bf c}$ which represents a goal node with $d_{H}(\hat{\bf c}_{0}^{k-1},\hat{\bf z}_{0}^{k-1})$ = $i$ will be searched before any path $\hat{\bf c}'$ which represents a goal node with $d_{H}(\hat{\bf c'}_{0}^{k-1},\hat{\bf z}_{0}^{k-1})$ = $i+1$ except for the the path $\hat{\bf c}''$ of which $d_{H}(\hat{\bf c''}_{0}^{k-1},\hat{\bf z}_{0}^{k-1})$ = $i+1$ where $\hat{c}''_{k-1} \neq \hat{z}_{k-1}$.

Proof : Let ${\bf e}_{a}^{b}(l)$ represent the vector $(e_a$, $e_{a+1}$, $e_{a+2}$, $...$, $e_{b})$, where $e_l$ = 1 and  $e_j = 0$ for $j \neq l$.  
A  node is generally represented by $\hat{\bf c}_{0}^{l}$ = $(\hat{c}_0$, $\hat{c}_{1}$, $...$, $\hat{c}_{l})$, $ 0 \le l \le n-1$.

In this proof of $A^*$ decoding using the modified stack, we waive the condition that a node should be eliminated and will not be in the stack if its metric is larger than $M_{best}$.  Such a waiver will not affect the order of search for all the goal nodes. 

The message parts of the first two goal nodes searched in the $A^*$ decoding are $\hat{\bf c}_{0}^{k-1}$ = $(\hat{z}_0, \hat{z}_1, ..., \hat{z}_{k-1})$ = $\hat{\bf z}_{0}^{k-1}$ and $\hat{\bf c}_{0}^{k-1}$ = $(\hat{z}_0, \hat{z}_1, ..., \hat{z}_{k-1}\oplus 1)$ = $\hat{\bf z}_{0}^{k-1}$ $+$ ${\bf e}_{0}^{k-1}(k-1)$ respectively.   
Hence, the $d_{H}(\hat{\bf c}_{0}^{k-1},\hat{\bf z}_{0}^{k-1})$ values of these two goal nodes are $0$ and $1$ respectively.
 Currently, the paths (nodes) in the stack are $\hat{\bf z}_{0}^{0}$ $+$ ${\bf e}_{0}^{0}(0)$, $\hat{\bf z}_{0}^{1}$ $+$ ${\bf e}_{0}^{1}(1)$, $\hat{\bf z}_{0}^{2}$ $+$ ${\bf e}_{0}^{2}(2)$, ..., $\hat{\bf z}_{0}^{k-2}$ $+$ ${\bf e}_{0}^{k-2}(k-2)$  respectively from top to the bottom, 
where $d_{H}(\hat{\bf c}_{0}^{l},\hat{\bf z}_{0}^{l})$ of each node $\hat{\bf c}_{0}^{l}$ = $\hat{\bf z}_{0}^{l}$ $+$ ${\bf e}_{0}^{l}(l)$ is 1 for $0 \le l \le k-2$.

From node $\hat{\bf z}_{0}^{0}$$+$${\bf e}_{0}^{0}(0)$ which is at the top of the stack, the $A^*$ decoding proceeds to the next two goal nodes for which the message parts are $\hat{\bf c}_{0}^{k-1}$ = $\hat{\bf z}_{0}^{k-1}$$+$${\bf e}_{0}^{k-1}(0)$ and  $\hat{\bf c}_{0}^{k-1}$ = $\hat{\bf z}_{0}^{k-1}$$+$${\bf e}_{0}^{k-1}(0)$$+$${\bf e}_{0}^{k-1}(k-1)$ respectively.  Hence, the $d_{H}(\hat{\bf c}_{0}^{k-1}, \hat{\bf z}_{0}^{k-1})$ values of these two goal nodes are 1 and 2 respectively.
Currently, nodes of the stack are sequentially
$\hat{\bf z}_{0}^{1}$$+$${\bf e}_{0}^{1}(1)$, $\hat{\bf z}_{0}^{2}+{\bf e}_{0}^{2}(2)$, ..., $\hat{\bf z}_{0}^{k-2}+{\bf e}_{0}^{k-2}(k-2)$, $\hat{\bf z}_{0}^{1}+{\bf e}_{0}^{1}(0)+{\bf e}_{0}^{1}(1)$, $\hat{\bf z}_{0}^{2}+{\bf e}_{0}^{2}(0)+{\bf e}_{0}^{2}(2)$, ..., $\hat{\bf z}_{0}^{k-2}$$+\hat{\bf e}_{0}^{k-2}(0)$$+{\bf e}_{0}^{k-2}(k-2)$ respectively. For each newly added node $\hat{\bf c}_{0}^{l}$ = $\hat{\bf z}_{0}^{l}+{\bf e}_{0}^{l}(0)+{\bf e}_{0}^{l}(l)$, the associated 
$d_{H}(\hat{\bf c}_{0}^{l},\hat{\bf z}_{0}^{l})$ value is 2, where $1 \le l \le k-2$.

In the following, for $1 \le l \le k-2$, $\hat{\bf z}_{0}^{l}+{\bf e}_{0}^{l}(l)$  is sequentially extended to two goal nodes for which the message parts are $\hat{\bf c}_{0}^{k-1}$ = $\hat{\bf z}_{0}^{k-1}$ + ${\bf e}_{0}^{k-1}(l)$ and  $\hat{\bf c}_{0}^{k-1}$ = $\hat{\bf z}_{0}^{k-1}$ + ${\bf e}_{0}^{k-1}(l)$ + ${\bf e}_{0}^{k-1}(k-1)$ respectively, where the $d_{H}(\hat{\bf c}_{0}^{k-1}, \hat{\bf z}_{0}^{k-1})$ values of these two goal nodes are 1 and 2 respectively. The new nodes added to the bottom of the stack are $\hat{\bf z}_{0}^{l+1}+{\bf e}_{0}^{l+1}(l)+{\bf e}_{0}^{l+1}(l+1)$, $\hat{\bf z}_{0}^{l+2}+{\bf e}_{0}^{l+2}(l)+{\bf e}_{0}^{l+2}(l+2)$, ..., $\hat{\bf z}_{0}^{k-2}+{\bf e}_{0}^{k-2}(l)+{\bf e}_{0}^{k-2}(k-2)$.  For each newly added node $\hat{\bf c}_{0}^{l'}$ = $\hat{\bf z}_{0}^{l'}+{\bf e}_{0}^{l'}(l'-1)+{\bf e}_{0}^{l'}(l')$, the associated 
$d_{H}(\hat{\bf c}_{0}^{l'},\hat{\bf z}_{0}^{l'})$ value is 2, where $l+1 \le l' \le k-2$.

By now, any goal node with message part $\hat{\bf c}_{0}^{k-1}$ where $d_{H}(\hat{\bf c}_{0}^{k-1},\hat{\bf z}_{0}^{k-1})$ = $1$ has been searched in the $A^*$ decoding using the modified stack.  In addition, any node $\hat{\bf c}_{0}^{l'}$ with $d_{H}(\hat{\bf c}_{0}^{l'},\hat{\bf z}_{0}^{l'})$ = $2$ which has not been searched is stored in the stack and in fact each node $\hat{\bf c}_{0}^{l'}$ in the stack has $d_{H}(\hat{\bf c}_{0}^{l'},\hat{\bf z}_{0}^{l'})$ = $2$.

Assume that each goal node with message part $\hat{\bf c}_{0}^{k-1}$ where  $d_{H}(\hat{\bf c}_{0}^{k-1},\hat{\bf z}_{0}^{k-1})$ = $\ell-1$ has been searched.  Also assume that each node $\hat{\bf c}_{0}^{l'}$ with $d_{H}(\hat{\bf c}_{0}^{l},\hat{\bf z}_{0}^{l'})$ = $\ell$ which has not been searched is stored in the stack, where $\ell = 1, 2, ..., k-2$.  Suppose that the node at the top of the stack is $\hat{u}_{0}^{f}$, where $d_{H}(\hat{\bf u}_{0}^{f},\hat{\bf z}_{0}^{f})$ = $\ell$.  The message parts of the two goal nodes extended from this top node are  $\hat{\bf u}_{0}^{k-1}$ =
$(\hat{u}_0, \hat{u}_1$, ..., $\hat{u}_{f}, \hat{z}_{f+1}, \hat{z}_{f+2}$, ..., $\hat{z}_{k-2}, \hat{z}_{k-1})$ and $\hat{\bf u}_{0}^{k-1}$$+\hat{\bf e}_{0}^{k-1}(k-1)$ = $(\hat{u}_0, \hat{u}_1$, ..., $\hat{u}_{f}, \hat{z}_{f+1}, \hat{z}_{f+2}$, ..., $\hat{z}_{k-2}, \hat{z}_{k-1}\oplus 1)$ respectively, where $d_{H}(\hat{\bf u}_{0}^{k-1},\hat{\bf z}_{0}^{k-1})$ = $\ell$ and $d_{H}(\hat{\bf u}_{0}^{k-1}$$+\hat{\bf e}_{0}^{k-1}(k-1),\hat{\bf z}_{0}^{k-1})$ = $\ell+1$. That means for each of the two goal nodes represented by $\hat{\bf c}_{0}^{n-1}$, the associated $d_{H}(\hat{\bf c}_{0}^{k-1},\hat{\bf z}_{0}^{k-1})$ is either $\ell$ or $\ell+1$.
The nodes added to the bottom of the stack are sequentially $\hat{\bf u}_{0}^{f+1}$$+\hat{\bf e}_{0}^{f+1}(f+1)$, $\hat{\bf u}_{0}^{f+2}$$+\hat{\bf e}_{0}^{f+2}(f+2)$, ..., $\hat{\bf u}_{0}^{k-2}$$+\hat{\bf e}_{0}^{k-2}(k-2)$, where the $d_{H}(\hat{\bf c}_{0}^{l'}, \hat{\bf z}_{0}^{l'})$ of each node is $\ell+1$.    Note that all the  nodes with $d_{H}(\hat{\bf c}_{0}^{l'}, \hat{\bf z}_{0}^{l'})$ = $\ell$ are on top of those newly added nodes with $d_{H}(\hat{\bf c}_{0}^{l'}, \hat{\bf z}_{0}^{l'})$ = $\ell+1$. 
By extending all the nodes in the stack with $d_{H}(\hat{\bf c}_{0}^{l'}, \hat{\bf z}_{0}^{l'})$ = $\ell$ to goal nodes, the nodes remained in the stack are all have $d_{H}(\hat{\bf c}_{0}^{l'}, \hat{\bf z}_{0}^{l'})$ = $\ell+1$.

$\square\square$

\end{theorem}

There is a disadvantage for the modified stack design.  Under the condition of a small stack size, using the modified stack design it is more likely to delete a maximum likelihood codeword as compared to using the conventional stack design.  Hence, a large stack size is needed for using the modified stack. In our simulation, the maximum number of nodes used in the modified stack is not greater than 60000.  For the modern IC technology, a stack size of 60000 is not a problem.

\subsection {Performances Regarding Different Stack Designs and Thresholds}
\begin{figure}[h]
  \centering
  \includegraphics[width=0.5\textwidth]{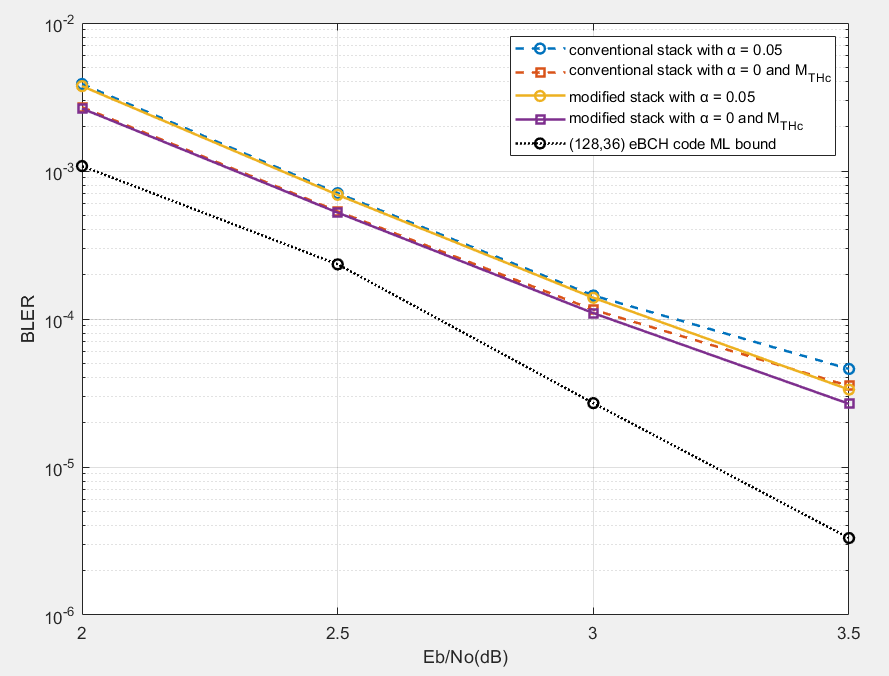}\\
  \caption{BLER performances for (128,36) eBCH code using PC-out-$\lambda$ with $\lambda$=4 based on modified stack with size of 60000 and conventional stack with size of 30000  respectively and using stopping criteria based on $\alpha$ = 0, $\alpha$ = 0.05 and $M_{TH,\hat{c}}$ respectively.}
  \label{eBCH128 36 BLER}
\end{figure}

\begin{figure}[h]
  \centering
  \includegraphics[width=0.5\textwidth]{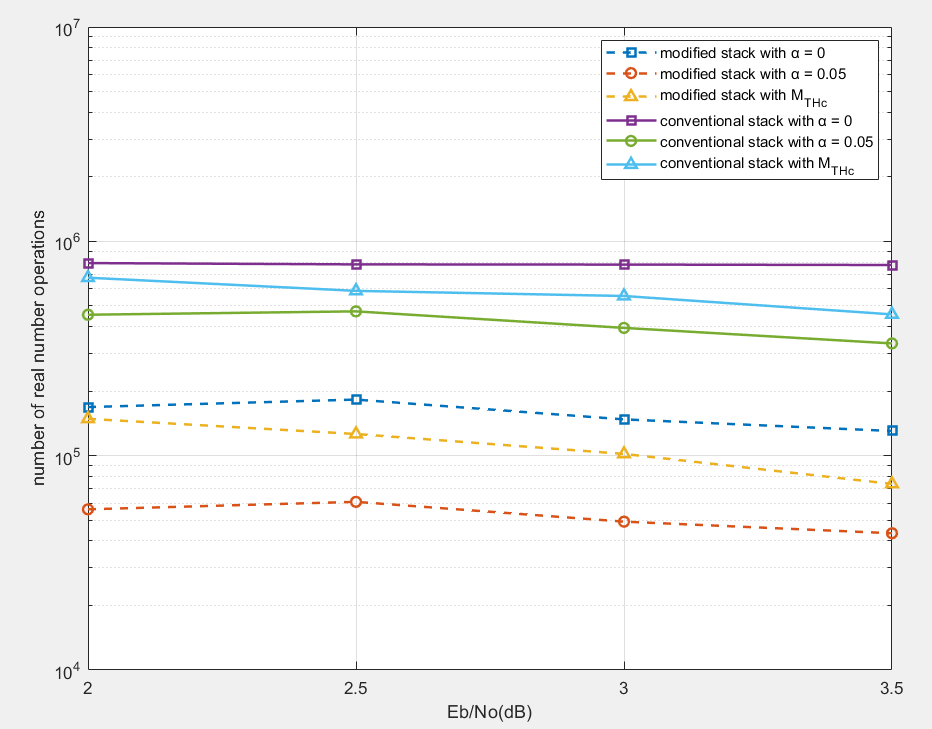}\\
  \caption{Number of real-number operations for (128,36) eBCH code using PC-out-$\lambda$ with $\lambda$=4 based on modified stack with size of 60000 and conventional stack with size of 30000 respectively and using stopping criteria based on $\alpha$ = 0, $\alpha$ = 0.05 and $M_{TH,\hat{c}}$ respectively.}
  \label{eBCH128 36 OP}
\end{figure}

For the comparison of conventional stack and modified stack, in Fig.~\ref{eBCH128 36 BLER} and Fig.~\ref{eBCH128 36 OP}, we show the BLER and the decoding complexity (i.e., the number of real number operations which is the sum of the number of visited edges per message bit and the number of comparisons per message bit) for the (128,36) extended BCH codes using the various stacking methods, where the thresholds include $M_{TH,\hat{c}}$ and $M_{TH,\alpha}$ with $\alpha$ = 0 and 0.05 respectively.  For the conventional stack, stack size is set as 30000.  For the modified stack, stack size is set as 60000. In the simulation,  PC-out-$\lambda$ with $\lambda$ = 4 is used.  We observe the following results :
\begin{itemize}
\item 
The modified stack with stack size of 60000 has BLER similar to the conventional stack with size of 30000, while the decoding complexity is much lower.
\item 
As compared to not using the stopping criterion (i.e., $\alpha$ = 0), using the stopping criterion with threshold of $M_{TH,\alpha}$ and $\alpha$ = 0.05 can achieve very close BLER performances  for $E_b/N_0$ no greater 3 dB and slightly higher BLER at $E_b/N_0$ = 3.5 dB.
\item 
As compared to not using the stopping criterion (i.e., $\alpha$ = 0), using the stopping criterion with $M_{TH,\hat{\bf c}}$ has the same BLER performances while the decoding complexity is lower.
\item 
As compared to using the stopping criterion with $M_{TH,\hat{\bf c}}$, using the stopping criterion with threshold $M_{TH,\alpha}$ of $\alpha = 0.05$ has slightly inferior BLER performances while the decoding complexity is lower.
\end{itemize}

\section{Short Codes Based on Concatenation}
To reduce the parameter "$\lambda$" of OSD-$\lambda$ or PC-out-$\lambda$ without sacrificing the error performances, we need to devise improved MRIP frame which is more reliable than the conventional MRIP frame.   We resort to short block codes, each of which is a concatenated code constructed by using a Reed-Solomon (RS) code as the outer code and a binary code as the inner code. 
Let $C_{out}$ be an ($n_{out},k_{out}$) Reed-Solomon code over $GF(2^m)$ and $C_{in}$ be an ($n_{in},k_{in}$) binary code, where $k_{in} = n_{out}m$.  The concatenated of $C_{out}$ and $C_{in}$ results in an ($n,k$) code, $C$ which is denoted as $C_{out} \circ C_{in}$, where $k = k_{out}m$ and $n_{in} = n$.  

\subsection{Possible Inner Codes}
The inner code needs to be capable of implementing efficient soft-in soft-out (SISO) decoding to provide reliable log-likelihood ratio (LLR) for each code bit.  Based on this requirement, we will consider the following constructions for inner codes.
\begin{enumerate}
  \item 
  $C_{in}$ is the cascade of $n/8$ (8,4,4) extended Hamming codes, where $n$ is a multiple of 8.
  \item 
  $C_{in}$ is the cascade of $n/16$ (16,8,5) binary block codes \cite{192220}, where $n$ is a multiple of 16.
  \item 
  $C_{in}$ is a (2,1,4) binary convolutional code without zero termination.
  \item 
  $C_{in}$ is a (2,1,6) binary convolutional code without zero termination.
  
\end{enumerate}

\subsection{Complexity of SISO decoding}
The SISO decoder for each of the (8,4,4) extended Hamming code and the (16,8,5) binary block code can be implemented by using brute force decoding, i.e., calculating all the codewords. There are $n/8$ (8,4,4) codes in $C_{in}$ and there are only 16 codewords in each (8,4,4) code. Hence, the complexity of implementing SISO decoding directly is very low. 

There are $2^8 = 256$ codewords in each (16,8,5) code and there are $n/16$ (16,8,5) inner codes for the (128,36) concatenated code $C$.  Hence, there are $(n/16)*256$ = 16$n$ inner codewords to calculate for the SISO decoding of $C_{in}$. The complexity for obtaining the LLR values for all the code bits in the improved MRIP of $C$ is about $(16n*16+128*8)/k$ real number operations per message bit, which is negligible as compared to the complexity of tree search for the $A^*$ decoder with PC-out-$\lambda$, $\lambda \ge 3$ (or the OSD decoder
with order $\lambda \ge 3$). 

For $C_{in}$ obtained from the either the (2,1,4) or the (2,1,6) convolutional code without termination, the SISO decoder can be easily implemented by the well known BCJR decoder \cite{1055186} with only one iteration.


\subsection{Improved MRIP Frame}
The $n$ LLR values for $C_{in}$ are represented by the vector ${\bf l}$ = $(l_0, l_1, ..., l_{n-1})$.  We propose an improved MRIP frame for which the role of ${\bf r}$ for the conventional MRIP frame is now replaced by ${\bf l}$. According to the magnitude of $|l_i|$, the vector ${\bf l}$ is permuted into the ${\bf l}'$ = $\Pi_{3}({\bf l})$ = $(l'_0, l'_1, ..., l'_{n-1})$, where magnitude of $|l'_i|$ is no less than $|l'_j|$ for $i < j$.   Further permutation $\Pi_{4}$ is applied to the columns of $\Pi_{3}(G)$ so that the first $k$ columns of $\Pi_{4}\Pi_{3}(G)$ are linearly independent, where the magnitudes of the first $k$ symbols of $\Pi_{4}\Pi_{3}({\bf l})$ are in descending order.  Then, $\Pi_{4}\Pi_{3}(G)$ is converted into a systematic generator matrix $\tilde{G}$, for which the first $k$ columns form an identity matrix.  Let $\tilde{\bf l}$ =  $\Pi_{4}\Pi_{3}({\bf l})$.  The hard decision of $\tilde{\bf l}$ is $\tilde{\bf z}$ = $(\tilde{z}_0, \tilde{z}_1,\cdots, \tilde{z}_{n-1})$, where $\tilde{z}_j = 0$ if $\mbox{sign}(\tilde{l}_j)$ = $1$ and $\tilde{z}_j$ = $1$ if $\mbox{sign}(\tilde{l}_j)$ = $-1$.  

For the $A^*$ decoding using the improved MRIP frame, permuted received vector $\tilde{\bf r}$ = $\Pi_{4}\Pi_{3}({\bf r})$, permuted hard decision vector $\tilde{\bf z}$ and the code tree generated by $\tilde{G}$ will be used, where each codeword in the code tree is  $\tilde{\bf c}$ = $(\tilde{c}_0, \tilde{c}_1, ..., \tilde{c}_{n-1})$.

\subsubsection{LLR of the SISO Output} 
To evaluate the potential of the improved MRIP frame obtained by a given inner code $C_{in}$, we will study the distribution of LLR for each code bit corresponding to the output of the SISO decoder based on each of the four possible inner codes provided in Sec. IV.A. 

 Let $r_i = (-1)^{c_{i}}$ + $w$, where $w$ is the additive Gaussian noise with variance $\sigma_{w}^2$.  The LLR value of $r_i$, i.e., the channel observation for $c_i$ is
\begin{equation}
\label{eqn: chobv}
L(r_i) = \ln\frac{p(r_i|c_i=0)}{p(r_i|c_i=1)} = \frac{2}{\sigma_{w}^2}((-1)^{c_{i}}+w).
\end{equation} 
It can be shown that the channel LLR, $L(r_i)$, is normally distributed, denoted $N(\mu_{L}(-1)^{c_{i}}$, $\sigma_{L}^2)$ for which its variance $\sigma_{L}^2 = 2\mu_{L}$ is twice of its mean $\mu_{L}$ = $2/(\sigma_w)^2$. Since $\sigma_{L}^2$ = $2\mu_{L}$,  the distribution of $L(r_i)$ can be easily represented by either its mean or its variance. 

 In \cite{957394}, it is suggested that the distribution denoted $L_i$ for the output LLR $l_i$ of SISO decoder for a block code can still be approximately modelled by $N(\mu_{L}(-1)^{c_{i}}$, $\sigma_{L}^2)$, where $\sigma_{L}^2$ is also approximated as twice of $\mu_{L}$.  

\begin{figure}[h]
  \centering
  \includegraphics[width=0.5\textwidth]{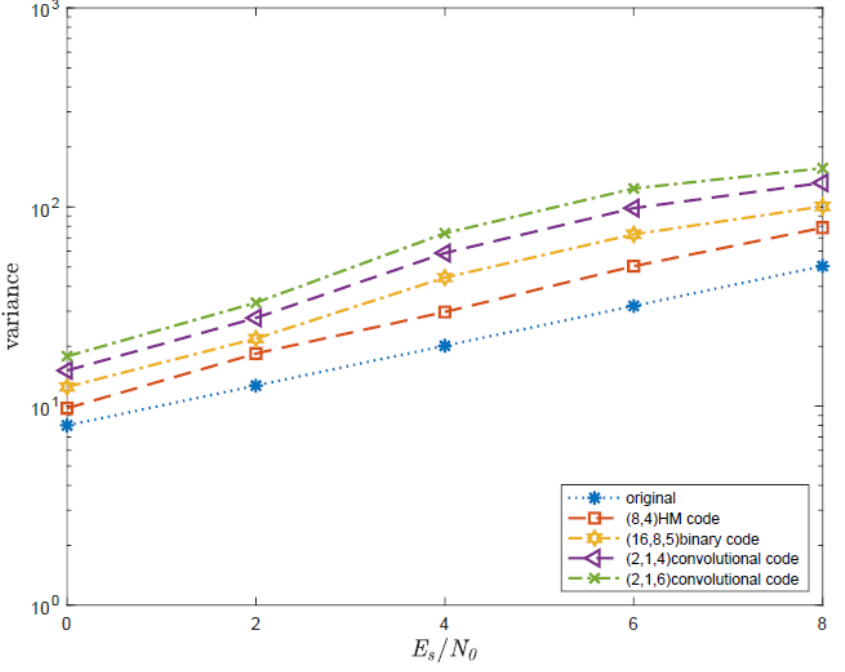}\\
  \caption{Variances $\sigma_{L}^2$ of LLR obtained from SISO decoders for various inner codes.}
  \label{variance-LLR}
\end{figure}

 The variances, $\sigma_{L}^2$, of LLR obtained from the SISO decoders for the four codes which will be used as inner codes of concatenated codes and the original channel observation shown in (\ref{eqn: chobv}) (i.e., no coding) are presented in Fig.~\ref{variance-LLR}, where the length $n$ of each of the (2,1,4) and (2,1,6) convolutional codes is set as 128.  For any $E_s/N_0$, the (2,1,6) binary code is the best and the original channel observation is the worst regarding the variance of LLR.  In the simulation for obtaining variance $\sigma_{L}^2$ for Fig.~\ref{variance-LLR}, we also obtain the mean $\mu_{L}$.  The simulated value of $\mu_{L}$ is indeed close to $\frac{1}{2}\sigma_{L}^2$.

\subsubsection{LLR of Each Code Bit After Ordering}
Assuming each $c_i = 0$, each $L_i$ in $(L_0, L_1, ..., L_{n-1})$ is now approximated by  $N(\mu_{L}$, $\sigma_{L}^2)$, where $\sigma_{L}^2$ = $2\mu_{L}$.  Applying ordered statistics to $(L_0, L_1, ..., L_{n-1})$, we have  $(L'_0, L'_1, ..., L'_{n-1})$, where the sampled value $l'_i$ of $L'_i$ is in decreasing order.

According the theory of ordered statistics, the probability density function of $L'_i$ is
\begin{equation} \label{equ:pdf}
f_{L'_i}(l) = \frac{n!}{(n-1-i)!(i)!}[F_{L}(l)]^{n-1-i}[1-F_{L}(l)]^{i}f_{L}(l),
\end{equation}
where $f_{L}(l)$ and $F_{L}(l)$ are respectively the probability density function and cumulative distribution function of the distribution $N(\mu_{L}$, $\sigma_{L}^2)$.  More specifically, we have
\begin{equation}
f_{L}(l) = \frac{1}{\sqrt{2{\pi}\sigma_{L}^2}}e^{\frac{(l-\frac{\sigma_{L}^{2}}{2})^2}{2\sigma_{L}^2}}
\end{equation}
 and 
 \begin{equation}
 F_{L}(l) = 1-\frac{1}{2}{\mbox{erfc}}(\frac{l-\frac{\sigma_{L}^{2}}{2}}{\sqrt{2}\sigma_{L}}),
 \end{equation}
where 
\begin{equation}
\mbox{erfc}(l) = \frac{2}{\sqrt{\pi}}\int_{l}^{\infty} e^{-t^2}dt
\end{equation}
With $f_{L'_i}(l)$, we can then calculate the expected value of $L'_{i}$, i.e., 
\begin{equation}
\mu_{L'_{i}} = \int_{-\infty}^{\infty} l f_{L'_i}(l) dl
\end{equation}

\begin{figure}[h]
  \centering
  \includegraphics[width=0.5\textwidth]{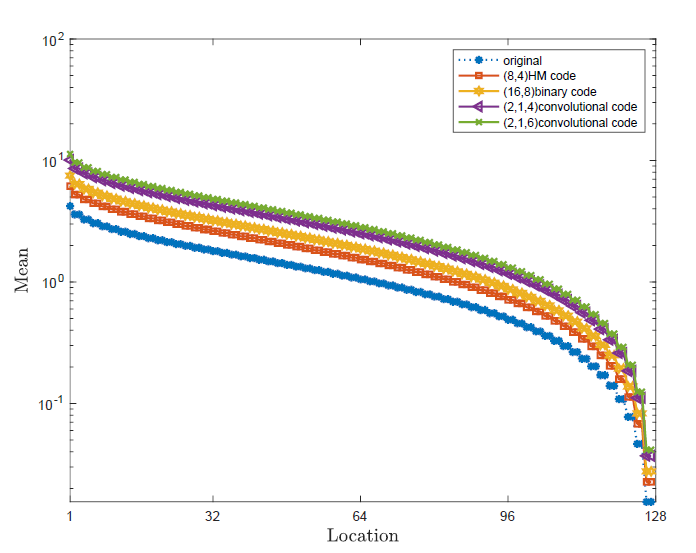}\\
  \caption{Means of LLR values in decreasing order for various inner codes. $E_s/N_0 = 3$ dB}
  \label{LLR decreasing}
\end{figure}

In Fig.~\ref{LLR decreasing}, we show $\mu_{L'_{i}}$, $i= 0, 1, 2, ..., 127$ for the four possible inner codes and the original channel observation, where $E_s/N_0 = 3$ dB and the values of $\mu_{L'_{i}}$ are obtained by simulation.  For each $i$, the four inner codes all provide $\mu_{L'_{i}}$ better than the original channel observation does.  Furthermore, among the four inner codes, the (2,1,6) convolutional code provides the best $\mu_{L'_{i}}$, the (2,1,4) convolutional code provides the second best, while the (8,4,4) has the worst $\mu_{L'_{i}}$.

\begin{figure}[h]
  \centering
  \includegraphics[width=0.5\textwidth]{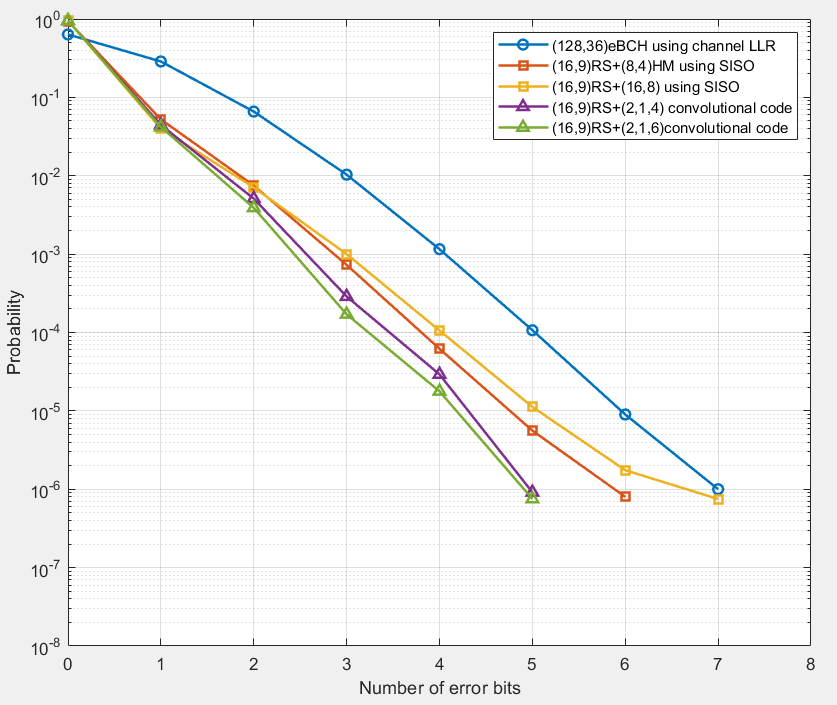}\\
  \caption{Probability regarding the number of errors at MRIP for (128,36) codes, $E_b/N_0$ = 3 dB. }
  \label{MRIP error}
\end{figure}

\begin{figure}[h]
  \centering
  \includegraphics[width=0.5\textwidth]{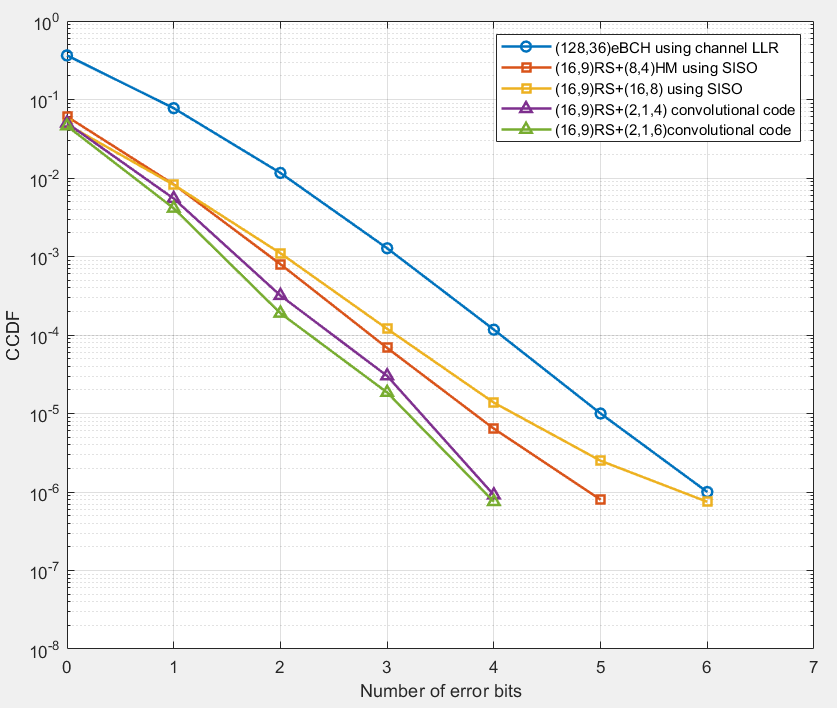}\\
  \caption{CCDF at MRIP for (128,36) codes, $E_b/N_0$ = 3 dB. }
  \label{CCDF}
\end{figure}

\begin{figure}[h]
  \centering
  \includegraphics[width=0.5\textwidth]{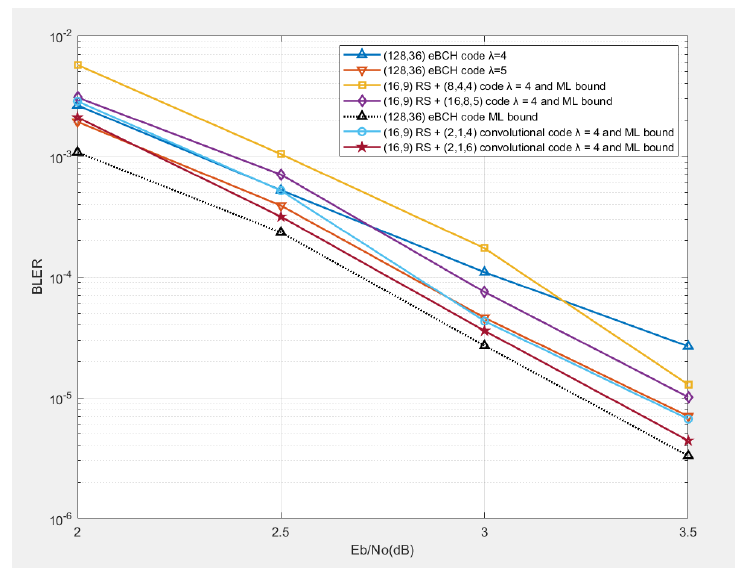}\\
  \caption{BLER for various (128,36) codes}
  \label{BLER 128 36}
\end{figure}

\section{Various Codes of Length Near 128}
In \cite{8594709}, it has been shown that the eBCH codes of length 128 using OSD with order 5, i.e., OSD-5 provide error rates better than turbo codes, LDPC codes and polar codes of length 128.  Hence, we use length-128 eBCH codes with various rates as benchmarks for comparison.  The  $A^*$ decoding is implemented using the modified stack with size of 60000 and using either $M_{TH,\hat{\bf c}}$ or $M_{TH,\alpha}$ with $\alpha = 0$.

\subsection{(128,36) codes}
We now compare the (128,36) eBCH code and several (128,36) binary codes constructed by using (16,9) Reed-Solomon code over GF(16) as the outer code $C_o$ and using various rate 1/2 binary inner codes $C_{in}$.  $A^*$ decoding with PC-out-$\lambda$, is used, where in most cases $\lambda$ = 4 are used.  In some cases, the simulation results with $\lambda$ = 5 are also provided.

\subsubsection{Errors at MRIP before $A^*$ Decoding} To see the advantage of the improved MRIP over the conventional MRIP regarding the reliability at the MRIP positions, we need to check $P(j|\mbox{MRIP})$, which is the probability of $j$ erroneous bits in the 36 MRIP bits in the hard decision vectors, i.e., $(\hat{z}_0, \hat{z}_1,\cdots, \hat{z}_{35})$ for the original case and $(\tilde{z}_0, \tilde{z}_1,\cdots, \tilde{z}_{35})$ for the concatenated coding case respectively.

By assuming $L'_i$, $i$ = 0, 1, 2, ..., 35, is normal distributed with mean $\mu_{L'_i}$ and variance $\sigma^{2}_{L'_i}$ = $2\mu_{L'_i}$, we can estimate $p_i$ which is the BER of $\tilde{z}_i$ by  the following equation,
\begin{equation}
p_i \approx \frac{1}{2} \mbox{erfc}(\frac{\sigma_{L'_i}}{2\sqrt{2}}) = \frac{1}{2} \mbox{erfc}(\frac{\sqrt{\mu'_i}}{2}).
\end{equation}
According to Fig.~\ref{variance-LLR}, we may estimate that, for any given $i$, using (2,1,6) convolutional code can yield the lowest $p_i$ while the original case yields the highest $p_i$.  However, it is difficult to calculate $P(j|\mbox{MRIP})$, $j = 0, 1, 2, ..., 5$ by using $p_i$, $i = 0, 1, ..., 127$.  The first reason is that $l'_0, l'_1, ..., l'_{35}$ may not necessarily be located at MRIP.  The second reason is that the random variable $L'_0, L'_1, ..., L'_{35}$ are not statistically independent. Hence, we resort to the direct simulation for obtaining $P(j|\mbox{MRIP})$.  The simulation results are provided in
Fig.~\ref{MRIP error}, where $E_b/N_0$ = 3 dB.
As expected, for $1 \le j \le 5$, using (2,1,6) convolutional code as the inner code can provide the lowest $P(j|\mbox{MRIP})$ and using the original LLR for the eBCH code yields the worst (highest) $P(j|\mbox{MRIP})$.  Note however that using the (8,4,4) code as the inner code can obtain lower $P(j|\mbox{MRIP})$ as compared to using the (16,8,5) inner code.

We then check the complementary cumulative distribution function (CCDF), i.e., 
\begin{equation}
P_{CCDF}(\lambda) = 1-\sum_{j=1}^{\lambda} P(j|\mbox{MRIP})
\end{equation}
The CCDF for these inner codes with $E_b/N_0$ = 3 dB are provided in Fig.~\ref{CCDF}.
We see that the trend of CCDF is similar to that of $P(j|\mbox{MRIP})$.  For each code $C$, $P_{CCDF}(\lambda)$ is a lower bound of BLER obtained by using the PC-out-$\lambda$.
From Fig.~\ref{CCDF}, we see that for the (128,36) eBCH code, its $P_{CCDF}(4)$ value is very close to its BLER shown in Fig.~\ref{eBCH128 36 BLER}, where modified stack with PC-out-4 and $M_{TH,\hat{\bf c}}$ is used.  



\subsubsection{BLER Using $A^*$ Decoding} The advantage of the improved MRIP over the conventional MRIP regarding its application to the $A^*$ decoding can be observed in Fig.~\ref{BLER 128 36}, which shows the BLER of the decoding, where either the threshold $M_{TH,\alpha}$ with $\alpha$ = 0 or the threshold $M_{TH,\hat{\bf c}}$ is used.  Note that the BLER performances of eBCH code using the conventional $A^*$ decoding with PC-out-$\lambda$, $\lambda$ = 4 are quite deviated from the eBCH (128,36) ML bound.  The minimum distance of (128,36) eBCH code is 32.  The value of $\min\{\lceil d_{min}/4 \rceil-1,k\}$ is 7.  Hence, using $\lambda$ = 4 is not enough.  If we increase $\lambda$ to 5, the BLER performances will be closer to the ML bound, where the BLER of eBCH code will be decreased to about $4.6 \times 10^{-5}$ for $E_{b}/N_{0}$ = 3 dB and $7 \times 10^{-6}$ for $E_{b}/N_{0}$ = 3.5 dB.  For comparison, the $P_{CCDF}(5)$ value of the (128,36) eBCH code is $10^{-5}$. 

Among the BLER results of various schemes shown in Fig.~\ref{BLER 128 36}, 
the concatenated coding using the (2,1,6) convolutional inner code and improved MRIP has the best BLER performances which are close to the the maximum likelihood (ML) bounds for the eBCH code for $E_{b}/N_{0}$ = 3 dB and 3.5 dB respectively.    

For the concatenated coding using the (8,4,4) inner codes, its BLER at $E_{b}/N_{0}$ = 3 dB is $1.7 \times 10^{-4}$ which is higher than its $P_{CCDF}(4)$ value.  This can be explained by the fact that for a correct codeword $\tilde {\bf c}$ with $d_{H}(\tilde {\bf c}_{0}^{k-1},\tilde {\bf z}_{0}^{k-1})$ $\le 4$ its correlation discrepancy metric $M(\tilde{\bf r},\tilde{\bf c})$ may be greater than $M(\tilde{\bf r},\tilde{\bf c'})$, where  $\tilde{\bf c'}$ is a codeword with $d_{H}(\tilde {\bf c'}_{0}^{k-1},\tilde {\bf z}_{0}^{k-1})$ $> 4$.  This argument also explains the phenomenon  that the concatenated code using a (16,8,5) code as the inner code has a worse $P_{CCDF}(4)$ value and a better BLER at $E_{b}/N_{0}$ = 3 dB as compared to the concatenated code using the (8,4,4) code as the inner code.

\subsubsection{ML Bound}
In the simulation, we also obtain the ML lower bounds for the four concatenated coding schemes using the $A^*$ decoding with the improved MRIP.  These ML lower bounds are extremely close to the simulated BLER and hence are not plotted in Fig.~\ref{BLER 128 36}.   These results implies that increasing the $\lambda$ parameter over 4 for the concatenated codes can barely improve the BLER performances. The ML bound is obtained by counting the number of codewords $\hat{\bf c}_{best}$ obtained in the decoding for which $M(\hat{\bf r}, \hat{\bf c}_{best})$ is greater than $M(\hat{\bf r}, \hat{\bf c})$, where $\hat{\bf c}$ is the correct codeword.


\subsubsection{Effect of Stopping Criterion}
As stated in Section II.E. and Section III.C., we see that using either the $M_{TH,{\bf c}}$ or $M_{TH,\alpha}$ for the stopping criterion can reduce the decoding complexity, while using $M_{TH,{\bf c}}$ will not affect the BLER and using $M_{TH,\alpha}$ may somewhat affect the BLER. The calculation of $M_{TH,{\bf c}}$ requires the knowledge of $d_{min}$ of $C$.  For $C$ constructed by concatenated coding using the convolutional inner code, its minimum distance $d_{min}$ is not available and hence we may use the threshold instead.  However, we observe that the loss of BLER may be small.  For example using the (2,1,6) inner code at $E_b/N_0$ = 3.5 dB, using $M_{TH,\alpha}$ with $\alpha$ = 0.05 will result in BLER of $4.9 \times 10^{-6}$ as compared to the BLER of $4.4 \times 10^{-6}$ which is obtained by using $\alpha$ = 0.

\subsection{(128,64) code and (130,65) code} 
\begin{figure}[h]
  \centering
  \includegraphics[width=0.5\textwidth]{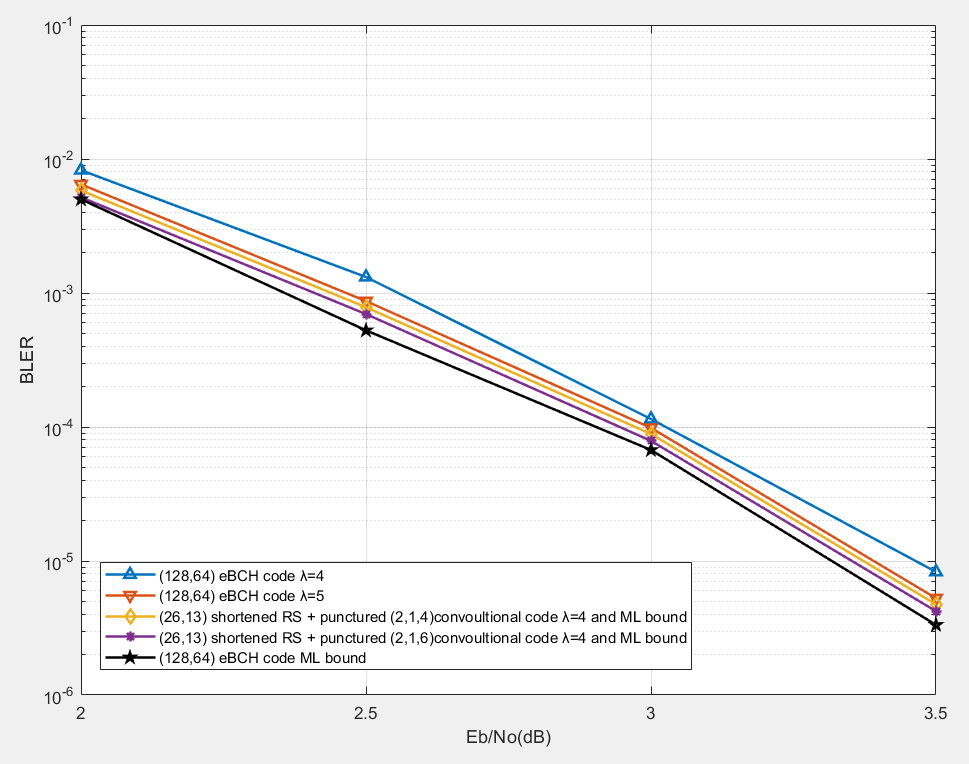}\\
  \caption{BLER for (128,64) and (130,65) codes }
  \label{BLER 128 64}
\end{figure}

The BLER performances of (128,64) eBCH code with $\lambda$ = 4 are provided in
Fig.~\ref{BLER 128 64}.  For comparison, we consider a (130,65) concatenated code which uses a (26,13) shortened Reed-Solomon code over GF($2^5$) as the outer code and a (2,1,4) (or (2,1,6)) convolutional code without termination as the inner code.  The BLER performances of eBCH code using the A* decoding with PC-out-$\lambda$ where $\lambda$ = 4 are not far from the eBCH (128,64) ML bound.  The minimum distance of (128,64) eBCH code is 22.  Hence, the value of $\min\{\lceil d_{min}/4 \rceil-1,k\}$ is 5.  
If we increase $\lambda$ to 5, the BLER of eBCH code will be a bit closer to the eBCH ML bound.  For 
the concatenated coding using either the (2,1,4) or the (2,1,6) convolutional inner code and improved MRIP $\lambda =4$ can achieve BLER performances which are almost identical to its  ML bound.   The (130,65) concatenated code using the (2,1,6) convolutional inner code $\lambda =4$ can achieve BLER performances slightly better than using the (128,64) eBCH code with $\lambda =5$.  

\begin{figure}[h]
  \centering
  \includegraphics[width=0.5\textwidth]{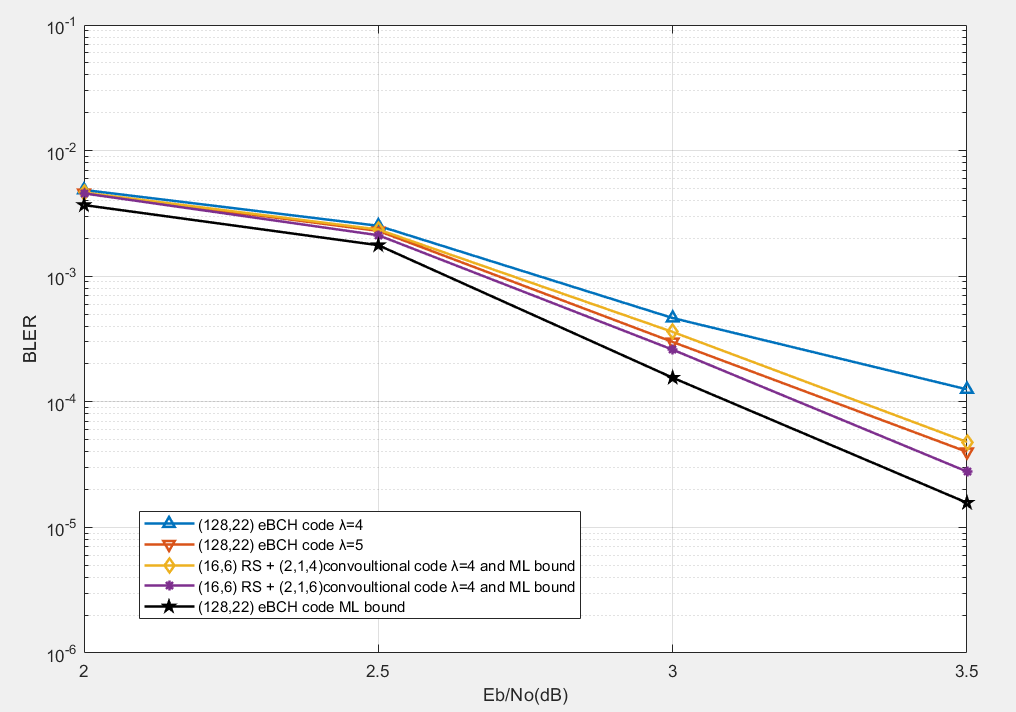}\\
  \caption{BLER for (128,22) codes }
  \label{BLER 128 22}
\end{figure}

\subsection{(128,22) code and (128,24) code}
The BLER performances of (128,22) eBCH code with $\lambda$ = 4 and $\lambda$ = 5 are provided in
Fig.~\ref{BLER 128 22}.  For comparison, we consider a (128,24) concatenated code which is constructed by using a (16,6) Reed-Solomon code over GF($2^4$) as the outer code and a (2,1,4) (or (2,1,6)) convolutional code without termination as the inner code.  The BLER performances of eBCH code using the A* decoding with PC-out-$\lambda$, $\lambda$ = 4 are far away from the eBCH (128,22) ML bound.  The minimum distance of (128,22) eBCH code is 48.  Hence, the value of $\min\{\lceil d_{min}/4 \rceil-1,k\}$ is 11.  
If we increase $\lambda$ to 5, the BLER of eBCH code will be improved but still is a bit distant from the eBCH ML  bound.  For 
the concatenated coding using either the (2,1,4) or the (2,1,6) convolutional inner code and improved MRIP $\lambda =4$ can achieve BLER performances which are extremely close to its ML bounds but not very close to the eBCH ML bound.  The (128,24) concatenated code using the (2,1,6) convolutional inner code $\lambda =4$ can achieve BLER performances better than using the (128,22) eBCH code with $\lambda =5$.

\subsection{Minimum Distance}
The minimum distance of the (128,36) concatenated code using (16,9) Reed-Solomon code over GF($2^4$) as the outer code and the (8,4,4) code as the inner code is $8 \times 4$ = 32.  The minimum distances of other concatenated codes are unknown.  Moreover, minimum distances of the (128,64), (128,36) , (128,22) eBCH codes are 22, 32 and 48 respectively.    We may examine the slopes of the ML bound curves to estimate the minimum distances of the concatenated codes with unknown minimum distances.  Asymptotically, the BLER is dominated by $\mbox{erfc}(\sqrt{\frac{d_{min}RE_{b}}{N_0}})$.  The factor $d_{min}R$ for the  (128,64), (128,36) , (128,22) eBCH codes are 11, 9 and 8.25 respectively.  Although our works are designed to operate in the low SNR region, we can tentatively  check the slopes for the three codes from $E_{b}/N_0$ of 3 dB to 3.5 dB.  The slope of ML bound  curve for for (128,64) eBCH code is steeper than those of the other two codes, while the slopes of the other two codes are similar.
For the (128,64), (130, 65), (128,22) concatenated codes using the (2,1,6) convolutional code as the inner code, the the slopes of ML bound from $E_{b}/N_0$ of 3 dB to 3.5 dB for the three codes from $E_{b}/N_0$ of 3 dB to 3.5 dB are similar to those for the comparable eBCH codes.  This observation suggests that their minimum distances are close to the comparable eBCH codes.  Note however that each of the three concatenated codes has the advantage of lower decoding complexity resulting from the improved MRIP frame obtained by SISO decoder for the inner code.

\section{Concluding Remarks}
In this research, we consider the construction and decoding of short binary code constructed by using a Reed-Solomon code as the outer code and a binary code as the inner code, where the soft-in soft-out (SISO) decoding is applied to the inner code.  The obtained  log-likelihood ratios are used to generate the improved MRIP, which enables us to obtain similar or better error performance with lower decoding complexity as compared to the decoding of eBCH cods of comparable length using the conventional MRIP.  In this paper, only codes with code length of 128 are considered. The idea of concatenated coding together improved MRIP can be easily applied to the construction and decoding of short codes with other code lengths.

The advantage of using the concatenated coding for the decoding of short codes are more evident for low-rate codes. 

\bibliographystyle{IEEEtran}

\bibliography{ref}

\begin{thebibliography}{10}
\providecommand{\url}[1]{#1}
\csname url@samestyle\endcsname
\providecommand{\newblock}{\relax}
\providecommand{\bibinfo}[2]{#2}
\providecommand{\BIBentrySTDinterwordspacing}{\spaceskip=0pt\relax}
\providecommand{\BIBentryALTinterwordstretchfactor}{4}
\providecommand{\BIBentryALTinterwordspacing}{\spaceskip=\fontdimen2\font plus
\BIBentryALTinterwordstretchfactor\fontdimen3\font minus
  \fontdimen4\font\relax}
\providecommand{\BIBforeignlanguage}[2]{{%
\expandafter\ifx\csname l@#1\endcsname\relax
\typeout{** WARNING: IEEEtran.bst: No hyphenation pattern has been}%
\typeout{** loaded for the language `#1'. Using the pattern for}%
\typeout{** the default language instead.}%
\else
\language=\csname l@#1\endcsname
\fi
#2}}
\providecommand{\BIBdecl}{\relax}
\BIBdecl

\bibitem{8594709}
M.~Shirvanimoghaddam, M.~S. Mohammadi, R.~Abbas, A.~Minja, C.~Yue, B.~Matuz,
  G.~Han, Z.~Lin, W.~Liu, Y.~Li, S.~Johnson, and B.~Vucetic, ``Short
  block-length codes for ultra-reliable low latency communications,''
  \emph{IEEE Communications Magazine}, vol.~57, no.~2, pp. 130--137, 2019.

\bibitem{412683}
M.~Fossorier and S.~Lin, ``Soft-decision decoding of linear block codes based
  on ordered statistics,'' \emph{IEEE Transactions on Information Theory},
  vol.~41, no.~5, pp. 1379--1396, 1995.

\bibitem{259636}
Y.~Han, C.~Hartmann, and C.-C. Chen, ``Efficient priority-first search
  maximum-likelihood soft-decision decoding of linear block codes,'' \emph{IEEE
  Transactions on Information Theory}, vol.~39, no.~5, pp. 1514--1523, 1993.

\bibitem{7182759}
T.-H. Chen, K.-C. Chen, M.-C. Lin, and C.-F. Chang, ``On $a^{\ast}$ algorithms
  for decoding short linear block codes,'' \emph{IEEE Transactions on
  Communications}, vol.~63, no.~10, pp. 3471--3481, 2015.

\bibitem{1291728}
A.~Valembois and M.~Fossorier, ``Box and match techniques applied to
  soft-decision decoding,'' \emph{IEEE Transactions on Information Theory},
  vol.~50, no.~5, pp. 796--810, 2004.

\bibitem{4039664}
W.~Jin and M.~P.~C. Fossorier, ``Reliability-based soft-decision decoding with
  multiple biases,'' \emph{IEEE Transactions on Information Theory}, vol.~53,
  no.~1, pp. 105--120, 2007.

\bibitem{1576590}
A.~Kothiyal, O.~Takeshita, W.~Jin, and M.~Fossorier, ``Iterative
  reliability-based decoding of linear block codes with adaptive belief
  propagation,'' \emph{IEEE Communications Letters}, vol.~9, no.~12, pp.
  1067--1069, 2005.

\bibitem{6952162}
C.-F. Chang, T.-Y. Lin, and M.-C. Lin, ``Tree-search decoding with path
  constraints for linear block codes,'' in \emph{2014 IEEE Wireless
  Communications and Networking Conference (WCNC)}, 2014, pp. 753--757.

\bibitem{8664355}
H.-S. Shih, C.-S. Wang, C.-C. Chen, and M.-C. Lin, ``Tree-search decoding using
  reduced-size stacks,'' in \emph{2018 International Symposium on Information
  Theory and Its Applications (ISITA)}, 2018, pp. 511--515.

\bibitem{192220}
T.~Gulliver and V.~Bhargava, ``A systematic (16,8) code for correcting double
  errors and detecting triple-adjacent errors,'' \emph{IEEE Transactions on
  Computers}, vol.~42, no.~1, pp. 109--112, 1993.

\bibitem{slin2004}
S.~Lin and D.~J. Costello, \emph{Error Control Coding}, 2nd~ed.\hskip 1em plus
  0.5em minus 0.4em\relax NJ, USA: Prentice-Hall, Inc., 2004.

\bibitem{arikan2019seq}
\BIBentryALTinterwordspacing
E.~Arikan, ``From sequential decoding to channel polarization and back again,''
  2019. [Online]. Available: \url{https://arxiv.org/abs/1908.09594}
\BIBentrySTDinterwordspacing

\bibitem{zhu23}
\BIBentryALTinterwordspacing
H.~Zhu, Z.~Cao, Y.~Zhao, D.~Li, and Y.~Yang, ``Fast list decoders for
  polarization-adjusted convolutional (pac) codes,'' \emph{IET Communications},
  vol.~17, no.~7, pp. 842--851, 2023. [Online]. Available:
  \url{https://ietresearch.onlinelibrary.wiley.com/doi/abs/10.1049/cmu2.12587}
\BIBentrySTDinterwordspacing

\bibitem{61132}
D.~Taipale and M.~Pursley, ``An improvement to generalized-minimum-distance
  decoding,'' \emph{IEEE Transactions on Information Theory}, vol.~37, no.~1,
  pp. 167--172, 1991.

\bibitem{1055186}
L.~Bahl, J.~Cocke, F.~Jelinek, and J.~Raviv, ``Optimal decoding of linear codes
  for minimizing symbol error rate (corresp.),'' \emph{IEEE Transactions on
  Information Theory}, vol.~20, no.~2, pp. 284--287, 1974.

\bibitem{957394}
S.~ten Brink, ``Convergence behavior of iteratively decoded parallel
  concatenated codes,'' \emph{IEEE Transactions on Communications}, vol.~49,
  no.~10, pp. 1727--1737, 2001.

\end{thebibliography}

\newpage

\vfill

\end{document}